\documentclass[12pt]{iopart}

\usepackage{graphicx}  
\begin{document}

\title[Predictions for the LHC: an Overview]{Predictions for the LHC: an Overview}

\author{N Armesto}

\address{Departamento de F\'{\i}sica de Part\'{\i}culas and IGFAE,\\
Universidade de Santiago de Compostela, 15782 Santiago de Compostela, Spain}
\ead{nestor@fpaxp1.usc.es}
\begin{abstract}
I present an overview of predictions for the heavy ion program at the Large Hadron Collider. It is mainly based on the material presented during the workshop 'Heavy Ion Collisions at the LHC - Last Call for Predictions', held in the frame of the CERN Theory Institute from May 14th to June 10th 2007. Predictions on both bulk properties and hard probes are reviewed.
\end{abstract}


\section{Introduction}

In this paper I present an overview of the predictions for Pb-Pb collisions at $\sqrt{s_{\rm NN}}=5.5$ TeV at the Large Hadron Collider, mainly based on the material presented during the workshop 'Heavy Ion Collisions at the LHC - Last Call for Predictions', held in the frame of the CERN Theory Institute from May 14th to June 10th 2007 \cite{Abreu:2007kv}. Such compilation (see also \cite{Borghini:2007ub}, and \cite{Bass:1999zq} for the analogous one for RHIC) should be useful for: (a) distinguishing pre- from post-dictions; (b) assuming that a model tested at RHIC (and eventually SPS) energies can be extrapolated to the LHC, the huge lever arm in energy provides very strong constraints; and (c) providing a frozen image of our present understanding of ultra-relativistic heavy ion collisions.

I have classified the predictions in: bulk properties (Section 2): multiplicities (see pre-RHIC predictions in \cite{Armesto:2000xh}), azimuthal asymmetries, hadronic flavor observables and correlations at low transverse momentum; and hard and electromagnetic probes (Section 3): high transverse momentum observables and jets, quarkonium and heavy quarks, and leptonic probes and photons, extensively documented in \cite{Accardi:2004be,Accardi:2004gp,Bedjidian:2004gd,Arleo:2004gn}. Unless otherwise stated, the predictions presented here can be found in \cite{Abreu:2007kv} and will be referenced by the name of their first author. Finally I draw some conclusions.

\section{Bulk properties}

\subsection{Multiplicities}

\begin{figure}[h]
\begin{center}
\includegraphics[height=11.3cm]{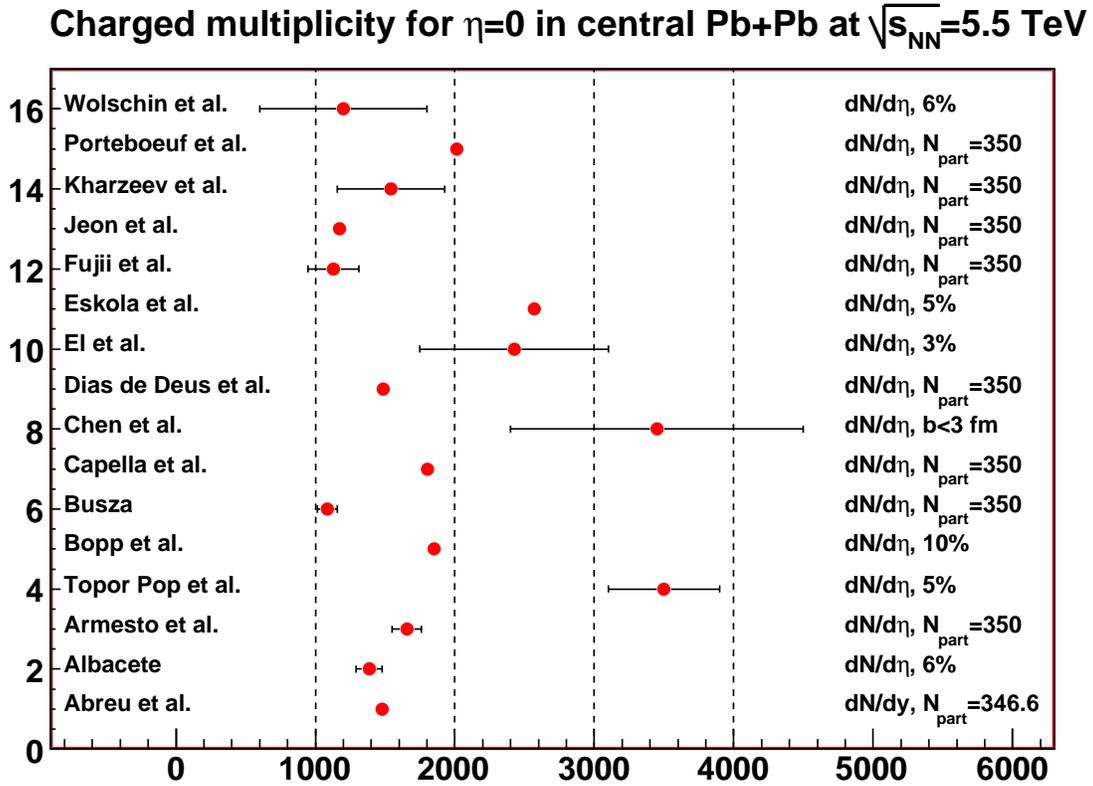}
\end{center}
\caption{Predictions for multiplicities in central Pb-Pb collisions at the LHC \cite{Abreu:2007kv}. On the left the name of the authors can be found. On the right, the observable and centrality definition is shown. The error bar in the points reflects the uncertainty in the prediction.}
\label{armestofig1}
\end{figure}

Multiplicity is a first-day observable. It plays a central role, as practically all other predictions require directly or indirectly such input. In Fig. \ref{armestofig1} sixteen predictions for charged hadron multiplicities at mid-rapidity in central collisions can be found.

\begin{figure}[h]
\begin{center}
\includegraphics[height=11.3cm]{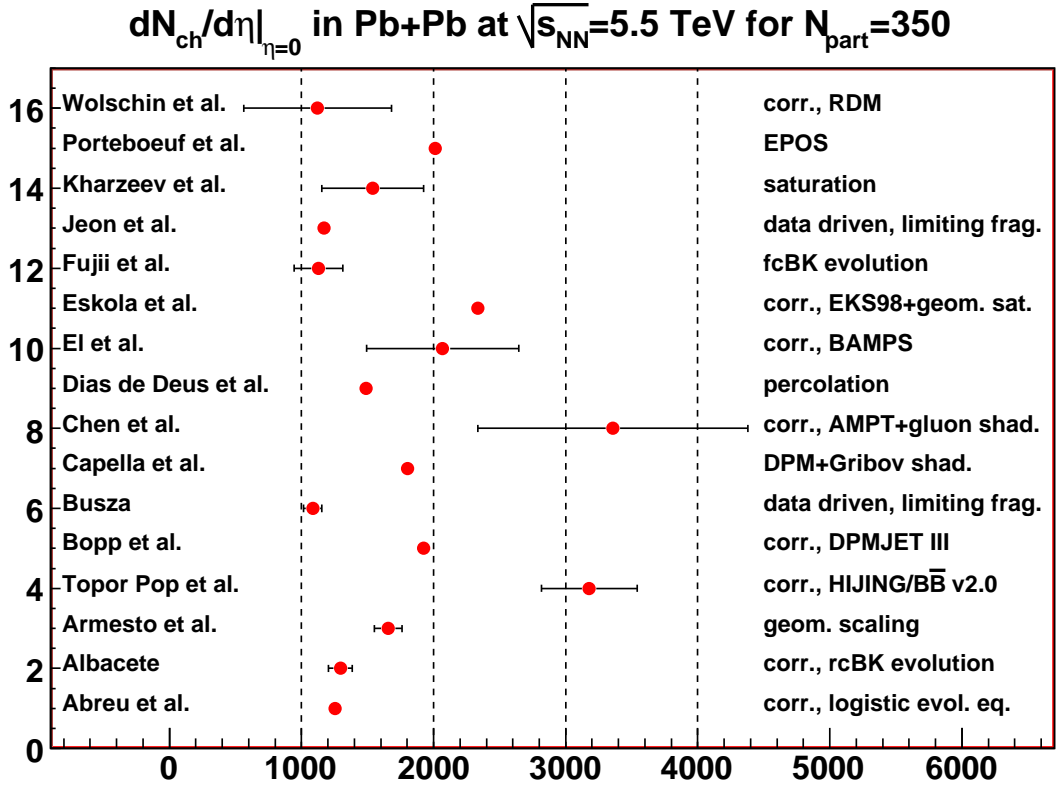}
\end{center}
\caption{Predictions for multiplicities in central Pb-Pb collisions at the LHC \cite{Abreu:2007kv}. On the left the name of the authors can be found. On the right, I indicate whether a correction has been applied or not, and provide a brief description of the key ingredients in the model. The error bar in the points reflects the uncertainty in the prediction.}
\label{armestofig2}
\end{figure}

To compare the different predictions more accurately, we re-scale them to a common observable ($dN_{ch}/d\eta|_{\eta=0}$) and centrality class ($\langle N_{part}\rangle =350$) using the model \cite{Amelin:2001sk}. The re-scaling factors can be read from Table \ref{armestotable1} and the corrected results found in Fig. \ref{armestofig2}. 
\begin{table}[h]
\caption{Results in the Monte Carlo code in  \cite{Amelin:2001sk} of the mean impact parameter, number of participants and binary nucleon-nucleon collisions, and charged multiplicity at mid-(pseudo-)rapidity, for different centrality classes defined by the number of participants.}
\label{armestotable1}
\begin{center}
\begin{tabular}{|c|c|c|c|c|c|}
\hline
\% & $ \langle b \rangle$ (fm) & $\langle N_{part} \rangle $ & $\langle N_{coll} \rangle$ & $dN_{ch}/dy|_{y=0}$ & $dN_{ch}/d\eta|_{\eta=0}$ \\ \hline
$0\div 3$ & 1.9 & 390 & 1584  & 3149 & 2633 \\ \hline
$0\div 5$ & 2.4 & 375 & 1490 & 2956 & 2472 \\ \hline
$0\div 6$ & 2.7 & 367 & 1447 & 2872 & 2402 \\ \hline
$0\div 7.5$ & 3.0 & 357 & 1390 & 2759 & 2306 \\ \hline
$0\div 8.5$ & 3.1 & 350 & 1354 & 2686 & 2245 \\ \hline
$0\div 9$ & 3.2 & 347 & 1336 & 2649 & 2214 \\ \hline
$0\div 10$ & 3.4  & 340 & 1303 & 2583 & 2159 \\ \hline
\end{tabular}
\end{center}
\end{table}

Predictions can be roughly classified into those based on saturation physics (Abreu et al., Albacete, Armesto et al., Eskola et al., Fujii et al., Kharzeev et al.), data-driven predictions (Busza, Jeon et al.), those based on percolation ideas (Dias de Deus et al.), those containing strong shadowing (Capella et al.), Monte Carlo models (Topor Pop et al., Bopp et al., Chen et al., Porteboeuf et al.) in which many physical mechanisms are combined, parton cascades (El et al.), and those based on diffusion equations (Wolschin et al.). They tend to lie in the range $1000\div 2000$, the lowest values corresponding to extreme saturation models and data-driven predictions. These values are generically much lower than pre-RHIC predictions  \cite{Armesto:2000xh,alicetp}. Most models contain now a large degree of collectivity in the form of saturation, strong gluon shadowing, percolation, etc.

\subsection{Azimuthal asymmetries}

Azimuthal asymmetries are another first-day observable.
Generically, $p_T$-integrated $v_2$ is expected to increase in all models. On the other hand, data-driven extrapolations \cite{Borghini:2007ub} tend to disagree with hydrodynamical models, and there are differences in the expectations for $v_2(p_T)$. Gathering the predictions in rough categories and focusing on the difference between $v_2$ at small $p_T\sim 1$ GeV/c at RHIC and at the LHC, we find:
\begin{itemize}
\item Hydrodynamical predictions: Drescher et al., on the basis of a simple model, predict deviations from ideal hydro seen at RHIC, to diminish at the LHC. Bluhm et al. Kestin et al. and Eskola et al. predict $v_2$ at small $p_T$ to be similar or slightly smaller than at RHIC on the basis of ideal hydrodynamics. Increasing viscous corrections would further diminish $v_2(p_T)$, although the initial conditions for evolution have to be settled.
\item Non-hydro predictions: Porteboeuf et al. find a similar $v_2(p_T)$ at the LHC than at RHIC, in the EPOS model. On the other hand, Chen et al. in the AMPT model, and the absorption model of Capella et al., show a moderate to strong increase. The parton cascade of Molnar indicates that at fixed viscosity-to-entropy ratio, $v_2(p_T)$ should decrease with increasing energy. This offers a possible  interpretation of the results in AMPT or absorption models in terms of an effective diminishing of the viscosity-to-entropy ratio due to the increasing parton/particle density.
\end{itemize}

\subsection{Hadronic flavor observables}

Three different groups of observables can be discussed:
\begin{itemize}
\item Particle ratios are usually studied in the framework of statistical hadronization models. Both Andronic et al. and Kraus et al. predict all antiparticle-to-particle ratios to be very close to 1 for the small $\mu_B\sim 1$ GeV expected at the LHC at mid-rapidity, with only $\bar{\rm p}$/p being slightly smaller. The latter group shows the sensitivity of multi-strange-to-non-strange hadron ratios on the temperature, and to deviations from the grand-canonical ensemble. Finally, Rafelski et al. show how non-equilibrium scenarios reflect on an increase of multi-strange hadron yields and a decrease of non-strange resonances, for several total multiplicity predictions. So different statistical hadronization scenarios could be distinguished. This becomes of capital importance in regeneration models for open and hidden heavy flavor production, as shown by the differences between the equilibrium (Andronic et al.) and the non-equilibrium (Kuznetsova et al.) predictions.
\item Net baryon number at mid-rapidity: p--$\bar{\rm p}$ is generically predicted $<4$ in a wide variety of models: those including the baryon junction mechanism like HIJING B/Bbar (Topor Pop et al.) or DPMJET (Bopp et al.), ideal hydro (Eskola et al.), EPOS (Porteboeuf et al.) and RDM (Wolschin et al.).
\item Baryon-to-meson ratios: Ideal hydro (Kestin et al.) and recombination (Chen et al.) models predicts larger values than models which consider the creation of large color fields (Topor Pop et al., Cunqueiro et al.). The latter predict the Cronin effect for protons to survive at the LHC. Both observables will clarify the hadronization and baryon-transport mechanisms.
\end{itemize}

\subsection{Correlations at low transverse momentum}

Concerning the HBT radii, all models (Chen et al. in AMPT, Kestin et al. with ideal hydro, the non-equilibrium scenarios of Sinyukov et al. and Karpenko et al., and the recent work \cite{Chojnacki:2007rq}) predict an increase, although the density (i.e. total multiplicity) dependence of such increase varies from model to model. Nevertheless, the predictive power is limited by the problems that the models face to reproduce RHIC data (i.e. the $k_T$-dependence of $R_{out}$ and $R_{side}$, and their relative magnitude), and the poor understanding of the role of dissipative effects.

\section{Hard and electromagnetic probes}

Before discussing the predictions, let me stress that the control of the benchmark for hard and electromagnetic probes will be at the LHC as key an aspect as it was at RHIC. For example, a basic ingredient as the extrapolation of nuclear parton densities to the LHC kinematical region, is under poor control. This can be seen e.g. in the factor $\sim 10$ uncertainty in the nuclear effects for the gluon distribution in Pb at $x\sim 10^{-4}$ that can be extracted from the comparison of the results shown in \cite{Accardi:2004be} and the recent work \cite{Eskola:2008ca}.

\subsection{High transverse momentum observables and jets}

\begin{figure}[h]
\begin{center}
\includegraphics[height=11.3cm]{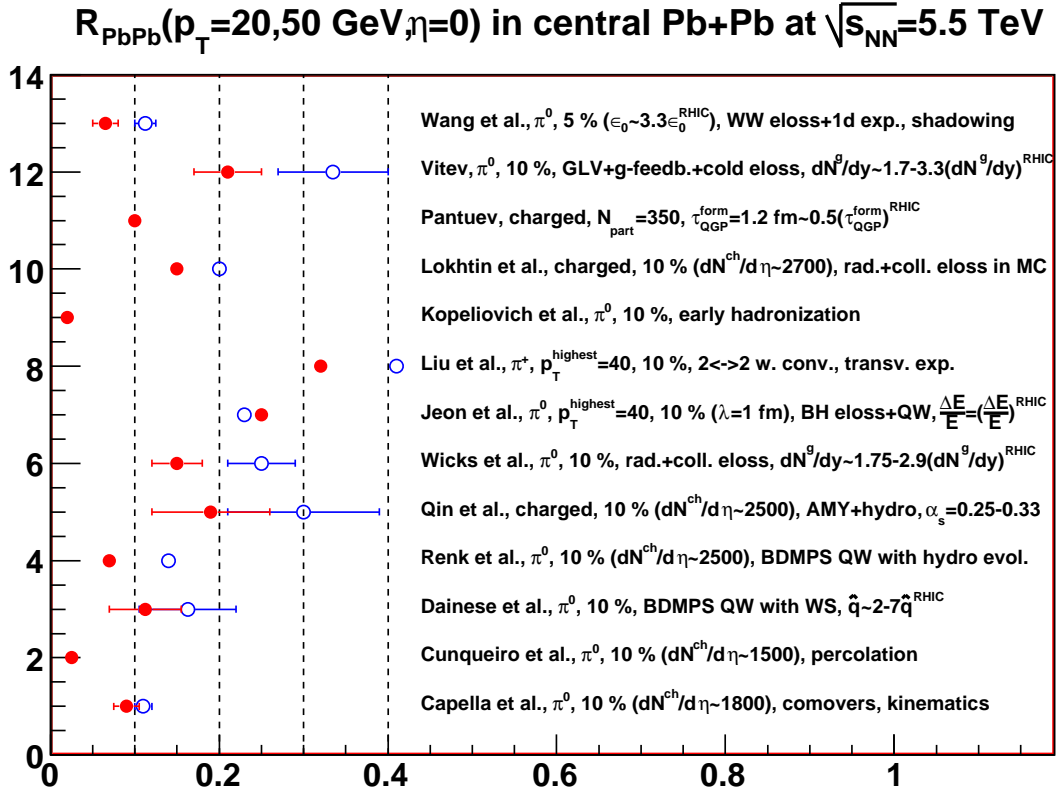}
\end{center}
\caption{Nuclear suppression factor $R_{\rm PbPb}(p_T,\eta=0)$ at $p_T=20$ (filled symbols) and 50 (open symbols) GeV/c from different models \cite{Abreu:2007kv}. On the right the authors of the predictions, the hadron species for which $R_{\rm PbPb}$ has been calculated, the centrality class, the density/multiplicity input, an indication in case the highest $p_T$ is 40 GeV/c instead of 50, and a brief description of the model ingredients, are provided. The error bars reflect the uncertainties coming from the multiplicity/density input but in Capella et al., Qin et al. and Wang et al.,  where they correspond to the variation of kinematics for particle production, $\alpha_s$ and nucleon/nuclear parton densities respectively.}
\label{armestofig3}
\end{figure}

In Fig. \ref{armestofig3}, I gather thirteen predictions for the nuclear suppression factor $R_{\rm PbPb}(p_T,\eta=0)$ in central collisions, for two values of $p_T=20$ and 50 GeV/c. Predictions can be roughly classified into those based on radiative and/or collisional energy loss in pQCD (Dainese et al., Renk et al., Qin et al., Wicks et al., Jeon et al., Liu et al., Lokhtin et al., Vitev, and Wang et al.), those based on absorption scenarios (Capella et al. and Kopeliovich et al.), those based on percolation (Cunqueiro et al.) and the opaque core-transparent corona model of Pantuev. The generic expectation in pQCD-based models is that $R^{\pi^0}_{\rm PbPb}(p_T=20 \, {\rm GeV/c},\eta=0)\simeq 0.1\div 0.2$ and increasing with increasing $p_T$.

Another aspect is the hadro-chemistry at large $p_T$, which in pQCD-based models is expected to be modified both by the different energy loss of quarks and gluons (Barnafoldi et al.), by the modification of the parton cascade inside the medium (Sapeta et al.), and by additional mechanisms as conversions (Liu et al.). Finally, jet reconstruction and other differential observables like the yield in the away-side peak (see predictions in Lokhtin et al. and Wang et al. respectively) will offer new possibilities to constraint models. All these aspects will help to verify the mechanism underlying the jet quenching phenomenon observed at RHIC.

\subsection{Quarkonium and heavy quarks}

Concerning heavy quarks, radiative energy loss models predict that the modification of the suppression pattern due to the mass of the quark will be visible in $R_{\rm AA}$ for heavy mesons and double ratios 
$R^B_{\rm AA}/R^D_{\rm AA}$ up to quite large $p_T\sim 20$ GeV/c (Armesto et al., Wicks et al.). These doubles ratios have also discriminatory power on the mechanism for energy loss (Horowitz et al.). A common feature to all pQCD-based mechanisms is that hadronization is supposed to happen outside the medium, which could not be the case for low $p_T$, where different mechanisms might be at work (van Hees et al., Vitev).

Concerning quarkonium suppression and production, the main uncertainty in recombination models for the magnitude of the suppression or enhancement lies in the heavy quark cross section (Andronic et al.). On the other hand, $p_T$-broadening as a tool to verify the recombination mechanism (Thews et al.) suffers from large uncertainties due to cold nuclear matter effects which should be clarified (Kang et al.) before any firm conclusion can be drawn.

The pattern of the suppression of the different quarkonium states versus $p_T$ should be measurable at the LHC. It offers a possibility to distinguish (Vogt) the different scenarios of sequential melting derived from either potential models or lattice computations, as well as new possibilities suggested in the framework of super-symmetric versions of QCD at strong coupling (H. Liu et al.). Finally, predictions which could be directly compared with experimental data, based on co-mover dissociation plus recombination, can be found in \cite{Capella:2007jv}. In any case, the considerable uncertainties in the mechanisms of quarkonium production and suppression limit our predictive power.

\subsection{Leptonic probes and photons}

Concerning photon production at large $p_T$, predictions in the frame of pQCD-based models containing radiative energy loss are available (Arleo, Vitev). At small $p_T$, other mechanisms as thermal production (Arleo et al.), conversions \cite{Liu:2008zb} or different types of factorization (Rezaeian et al.) could be at work. The measurement of $v_2$ for photons also offers some discriminatory power. Ideal hydro models (Chatterjee et al.) indicate that $v_2$ below $\sim 2$ GeV/c should be dominated by the partonic phase of the expansion of the medium. At the LHC, both high- and low-$p_T$ photons will be measurable, but an accurate understanding of the benchmark will be key to disentangle a thermal contribution over the other possible, background effects.

Concerning dileptons, Nayak et al. propose the ratio real-to-virtual photons as sensitive to the temperature of the initial phase and weakly dependent on the details of expansion, EOS, etc. Dremin proposed the Cherenkov effect as a mechanism which could mimic an apparent broadening of the $\rho$. Finally, predictions for the dilepton mass spectrum for both large- (Fries et al.) and small-$p_T$ (van Hees et al.) regions, are available. The intermediate mass region $1<M<3$ GeV/c$^2$ suffers from a large background from heavy meson semi-leptonic decays. Therefore the understanding of the benchmark looks crucial again to disentangle a thermal component.

\section{Conclusions}

We are getting closer and closer to the moment when RHIC-tested models will be confronted to LHC experimental results. For every observation at RHIC which is considered one of its major discoveries, an interpretation has been proposed and the corresponding predictions for the LHC have been computed. In this way:
\begin{itemize}
\item The charged multiplicity at mid-rapidity, sizably lower than expectations \cite{Bass:1999zq,Armesto:2000xh}, is interpreted in terms of a large degree of coherence or collectivity in particle production. It leads to predictions for central collisions at the LHC smaller than 2000 charged particles per unit pseudo-rapidity at $\eta=0$.
\item $v_2$ in agreement with ideal hydrodynamical models is interpreted as the creation of an almost perfect fluid, leading to the prediction that $v_2$ at small $p_T$ will be at the LHC similar or smaller than at RHIC.
\item The observed strong jet quenching leads to the conclusions that the created medium is dense, partonic and opaque, resulting in an $R^{\pi^0}_{\rm PbPb}(p_T=20 \, {\rm GeV/c},\eta=0)\simeq 0.1\div 0.2$ for central collisions at the LHC.
\end{itemize}

This picture has already motivated many theoretical developments: applications of AdS/CFT, mechanisms for early thermalization, viscous hydro, the CGC, etc. Major deviations from the generic expectations outlined here will enlarge our understanding of the Physics of ultra-relativistic heavy ion collisions - note that naive, data-driven extrapolations \cite{Borghini:2007ub} tend to disagree with predictions from successful models at RHIC. Besides, the LHC experiments, with their unprecedented capabilities \cite{Carminati:2004fp,Alessandro:2006yt,D'Enterria:2007xr,Steinberg:2007nm}, will provide new observables like jets or higher quarkonium states - not or only marginally measurable at RHIC - for which a large theoretical effort is still required.

\ack
This work has been supported by MEC of Spain under a Ram\'on y Cajal contract and grant FPA2005-01963, by the Spanish Consolider-Ingenio 2010 Programme CPAN (CSD2007-00042), and by Xunta de Galicia (Conseller\'{\i}a de Educaci\'on and grant PGIDIT07PXIB206126PR).
I thank the organizers for such a nice conference, J. Albacete, F. Bopp, W. Busza, L. Cunqueiro, A. Dainese, A. El, K. Eskola, U. Heinz, C.-M. Ko, I. Lokhtin, G. Milhano, C. Pajares, V. Pantuev, T. Renk, V. Topor Pop, R. Venugopalan, I. Vitev, X. N. Wang, K. Werner and G. Wolschin for feedback on their predictions, and J.  Albacete, J. Casalderrey-Solana, K. Eskola, E. Ferreiro, U. Heinz, P. Jacobs, X. N. Wang and U. Wiedemann  for discussions. Special thanks go to  N. Borghini, S. Jeon and U. A. Wiedemann who co-organized together with me the CERN workshop, and to C. Salgado for discussions and a critical reading of this manuscript.

\end{document}